\documentclass[aps,prl,twocolumn,floatfix,showpacs]{revtex4}
\def\gtwid{\mathrel{\raise.3ex\hbox{$>$\kern-.75em\lower1ex\hbox{$\sim$}}}}
\def\ltwid{\mathrel{\raise.3ex\hbox{$<$\kern-.75em\lower1ex\hbox{$\sim$}}}}

\usepackage{graphics}
\usepackage{epsfig}

\begin{document}
\title{Further Evidence for Neutrino Flux Variability from Super-Kamiokande
Data}
\author{D.O. Caldwell}
\affiliation{Physics Department, University of California, Santa Barbara,
                CA 93106-9530, USA}
\author{P.A.~Sturrock}
\affiliation{Center for Space Science
        and Astrophysics, Stanford University, Stanford, CA 94305-4060, USA}
\date{\today}

\begin{abstract}
While KamLAND apparently rules out Resonant-Spin-Flavor-Precession (RSFP)
as an explanation of the solar neutrino deficit, the solar neutrino fluxes
in the Cl and Ga experiments vary with solar rotation rates.  Added to this
evidence, summarized here, a power spectrum analysis of the Super-Kamiokande
data reveals (99.9\% CL) an oscillation in the band of twice the equatorial
rotation frequencies of the solar interior.  An $m=2$ magnetic structure
and RSFP, perhaps as a subdominant process, would give this effect.  Solar
cycle data changes are seen, as expected for convection zone modulations.
\end{abstract}
\maketitle

% body of paper here

Recent results from the KamLAND experiment \cite{ref:1} seem to confirm the Large-Mixing Angle (LMA)
solution to the solar neutrino deficit and rule out the
Resonant-Spin-Flavor-Precession (RSFP) explanation \cite{ref:2}.
On the other hand, there is increasing evidence \cite{ref:3}-\cite{ref:9}
that the solar neutrino flux is not constant as assumed for the LMA solution,
but rather it varies with well known solar rotation periods.  The solar
neutrino situation may be complex, and Spin-Flavor-Precession could
be subdominant to LMA, as suggested in \cite{ref:10}, or if there is at
least one light sterile neutrino even RSFP could be a subdominant process.
Since this recent information on solar neutrino variability is not widely known,
a brief summary is presented of analyses of radiochemical neutrino data, along with new
input from the Super-Kamiokande experiment \cite{ref:11}.  While the
10-day averages of Super-Kamiokande solar neutrino data
\cite{ref:12} show no obvious time dependence, a power-spectrum analysis \cite{ref:13a}
displays a strong peak at the frequency $26.57\pm0.05$
y$^{-1}$ (period 13.75 d), where the width of peaks are computed for a probability
drop to 10\%.  The probability of finding this peak (or a
stronger peak) by chance within a band specified by twice the
near-equatorial rotation frequencies of the solar interior (25.36-27.66 y$^{-1}$) is found
to be 0.001 (99.9\% CL).  This frequency is
also seen in the radiochemical data.

Whether or not it is the correct interpretation, the RSFP framework provides
a simple way to understand the data, and hence it will be used here.  The RSFP mechanism, which
requires a neutrino transition magnetic moment, provides at least as
good a global fit \cite{ref:13,ref:14} to solar data (which depends
mainly on the Super-Kamiokande spectrum) and a better fit to average rates
of the individual experiments than does LMA.  This is because
the neutrino survival probability, while having a resonance pit at a
density that suppresses the 0.86 MeV $^7$Be line (as does the
Small-Mixing-Angle (SMA) solution), tends at high energies
toward 1/2 and hence fits the Super-Kamiokande spectrum, whereas the
survival probability goes to unity in the SMA case.  Thus,
as a subdominant process, it could also improve fits to the data.

\begin{figure}
\includegraphics{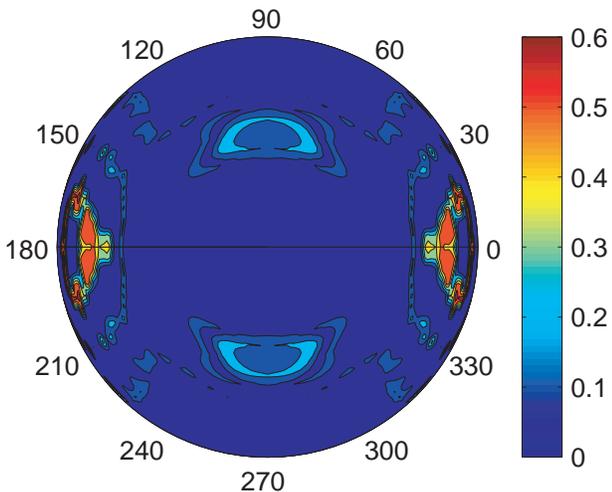}
\caption{Map of the resonance statistic of the SOHO/MDI helioseismology
and GALLEX data on a meridional section of the solar interior.  The only
high probability areas (red) are lens-shaped sections near the equator,
and all others are low probability.
\label{fig:1}}
\end{figure}

\begin{figure}
\epsfig{file=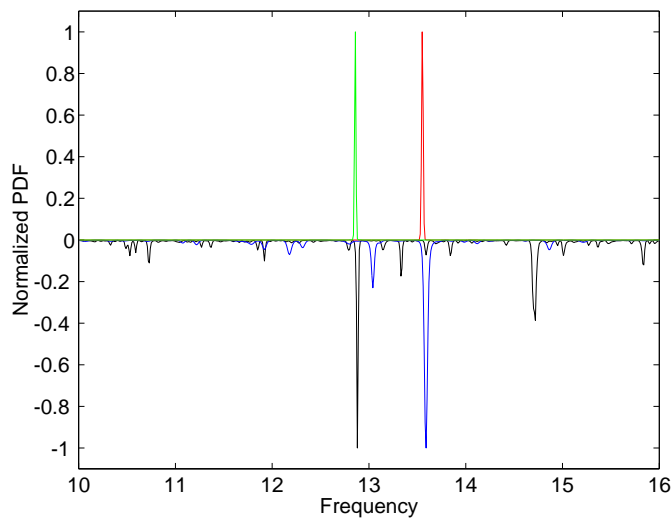,width=3.5in}
\caption{Comparison of normalized probability distribution functions formed
from power spectra of data from SXT equator (red), SXT N60-S60 (green),
Homestake (black), and GALLEX (blue).  Note that the SXT (red) and
GALLEX data are equatorial, and the other two are not.
\label{fig:2}}
\end{figure}

A brief review follows of the published evidence for solar neutrino flux
variability, put into a coherent neutrino scheme.  By analyzing $10^3$ simulated
data sequences, it was found \cite{ref:3} that the variance of the Homestake
solar neutrino data \cite{ref:15} is larger than expected at the 99.9\% CL.
A power spectrum analysis \cite{ref:3} of the data showed a peak at
$12.88\pm0.02$ y$^{-1}$ (28.4 d), compatible with the rotation rate of the solar radiative
zone.  Four sidebands gave evidence at the 99.8\% CL for a latitudinal
effect associated with the tilt of the sun's rotation axis, and the latitude
dependence was also seen directly in the data at the 98\% CL \cite{ref:4}.
The GALLEX data \cite{ref:16} showed a peak at $13.59\pm 0.04$ y$^{-1}$, compatible with the
equatorial rotation rate of the deep convection zone.  This peak is also in the
Homestake data, and a combined analysis of both data sets shows that the 13.59 y$^{-1}$
peak is larger than in either data set alone.  Comparison \cite{ref:7} of the power
spectrum for the GALLEX data with a probability distribution function for
the synodic rotation frequency as a
function of radius and latitude, derived from SOHO/MDI helioseismology data
\cite{ref:17}, results in Fig.~\ref{fig:1}.  This map shows the rotation
frequency coincides with the neutrino modulation in the equatorial section
of the convection zone at about 0.8 of the solar radius, $R_\odot$.
The influence of these rotation frequencies extends even to the corona,
since the SXT instrument \cite{ref:18} on the Yohkoh spacecraft provides
X-ray evidence for two ``rigid'' rotation rates, one ($13.55\pm0.02$ y$^{-1}$)
mainly at the equator, and the other ($12.86\pm0.02$ y$^{-1}$) mainly at high
latitudes.  These values are in remarkable agreement \cite{ref:8} with the neutrino modulation
frequencies and also with their equatorial location in one case and non-equatorial
in the other; see Fig.~\ref{fig:2}.

That two dominant frequencies are shown by both coronal X-rays and neutrino
flux is a feature of magnetic rotation well known in solar physics.  For instance, an
analysis \cite{ref:19} of the photospheric magnetic field during solar cycle
21 found two dominant magnetic regions: one in the northern hemisphere with
synodic rotation frequency $\sim13.6$ y$^{-1}$, and the other in the southern
hemisphere with synodic rotation frequency $\sim13.0$ y$^{-1}$.  Similarly, an
analysis \cite{ref:20} of flares during solar cycle 23 found a dominant
synodic frequency of $\sim13.5$ y$^{-1}$ for the northern hemisphere and $\sim12.9$
y$^{-1}$ for the southern hemisphere.  These and other studies show a strong
tendency for magnetic structure to rotate either at about 12.9 y$^{-1}$ or
at about 13.6 y$^{-1}$.  Since the sun's magnetic field is believed to
originate in a dynamo process at or near the tachocline, it is possible that
some magnetic flux is anchored in the radiative zone just below the tachocline,
where the synodic rotation frequency is about 12.9 y$^{-1}$, and some just
above the tachocline in the convection zone, where the synodic rotation
frequency is about 13.6 y$^{-1}$.  It is also possible that the 12.9 y$^{-1}$
frequency results from a latitudinal wave motion in the convection zone excited
by structures at or near the tachocline.

An example of latitudinal oscillatory motion of magnetic structures may be the
well-known Rieger-type oscillations with frequencies of about 2.4, 4.7, and 7.1 y$^{-1}$
\cite{ref:21}.  These periodicities may be attributed to r-mode oscillations with
spherical harmonic indices $\ell=3$, $m=1$,2,3, giving frequencies
$\nu=2m\nu_R/\ell(\ell+1)$ which are seen in neutrino data \cite{ref:3,ref:5,ref:9}.
A joint spectrum analysis \cite{ref:9} of Homestake and GALLEX-GNO data yields
peaks at 12.88, 2.33, 4.62, and 6.94 y$^{-1}$, indicating a sidereal rotation frequency
$\nu_R = 13.88\pm 0.03$ y$^{-1}$.  While $l\geq 2$ is required,
odd$-l$ values have nonzero poloidal velocity at the equator
that could move magnetic regions in or out of the neutrino
beam to earth.

The difference between the main modulations detected by the Cl and Ga
experiments may be explained by the tilt of the solar axis relative
to the ecliptic, along with the fact that Cl and Ga
neutrinos are produced mainly at quite different radii.  The Ga data comes
mostly---especially as the fit requires suppressing the $^7$Be line---from
$pp$ neutrinos, which originate predominantly at large solar radius
($\sim0.2$ R$_\odot$), so that the wide beam of neutrinos detected on earth
is insensitive to axis tilt.  Thus the beam of neutrinos detected by the Ga
experiments exhibits no seasonal variation and can be modulated by the equatorial
structure indicated in Fig.~\ref{fig:1}, leading to the observed frequency
at about 13.6 y$^{-1}$.
On the other hand, the Homestake experiment detects neutrinos produced from a
smaller sphere ($\sim0.05$ R$_\odot$), so that the axis tilt causes these
neutrinos mainly to miss the equatorial structure of Fig.~\ref{fig:1} and
instead sample nonzero latitudes where the
12.9 y$^{-1}$ modulation may occur.  Twice a year axis tilt has no effect for
these neutrinos, leading to a seasonal variation in the measured flux
\cite{ref:4}.

The 13.6 y$^{-1}$ frequency, located as in Fig.~\ref{fig:1}, represents a
modulation of the $pp$ neutrinos which are at or near the steeply falling edge of the neutrino
survival probability.  This is close to the RSFP resonance pit suppressing the $^7 \rm{Be}$ neutrinos, which is where
the largest value of $\Delta m^2/E$ satisfies
\begin{equation}
\Delta m^2/E=2\sqrt 2G_FN_{\rm eff},
\label{eq:1}
\end{equation}
and is essentially the same as for an MSW resonance \cite{ref:23}.  For MSW,
$N_{\rm eff}=N_e$ (the electron
density), and for RSFP $N_{\rm eff}=N_e-N_n$ for Majorana neutrinos, where
$N_n$ is the neutron density (about $N_e/6$ in the region of interest
in the Sun).  For Dirac neutrinos $N_{\rm eff}=N_e-N_n/2$, but only Majorana
neutrinos provide a fit to solar experimental rates with RSFP
\cite{ref:13,ref:14}, a very important consequence of such a solar solution.
Using the $N_e$ and $N_n$ values at the 0.8 R$_\odot$ location of
Fig.~\ref{fig:1} in Eq.~\ref{eq:1} results in $\Delta m^2/E\sim10^{-14}$ eV,
putting the RSFP resonance pit at the location needed to fit the solar data. $N_{\rm eff}$ varies exponentially with radius and would be quite
different for a somewhat changed neutrino modulation frequency.

\begin{figure}
\includegraphics{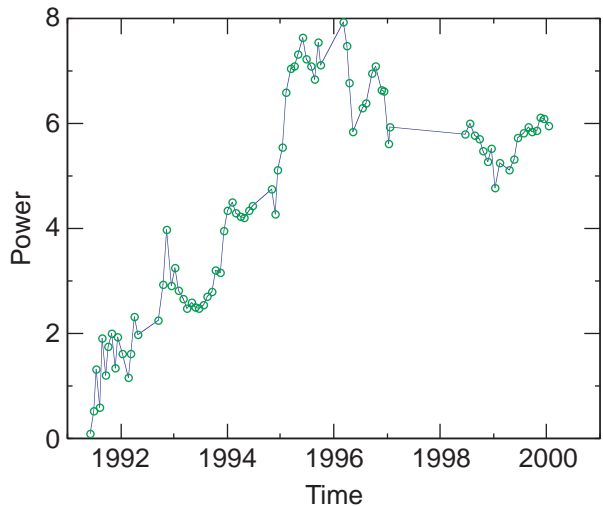}
\caption{Cumulative Rayleigh power vs. time for the 13.59 y$^{-1}$ frequency
peak from the GALLEX-GNO data.  Note that the power builds up from the start
of data taking after the 1990 solar maximum until the 1996.8 solar minimum,
after which there is little evidence for that frequency.
\label{fig:3}}
\end{figure}

It is well known that the convection zone magnetic field changes with the solar
cycle, so the neutrino modulation features described above are not permanent.
It has been argued \cite{ref:24} that an RSFP effect would have to be in the
radiative zone, where the field does not change with the solar cycle, under
the assumption solar
cycle variations are not observed.  On the contrary, we show here that solar
cycle changes play an important role.  Variations in neutrino rates are difficult to
observe, since changes in field magnitude would be undetectable if the transition
remains adiabatic, and even if flux modulation results, average rates may vary
only slightly.
The feature that is most sensitive to field magnitude or radial variations
is the intersection of the very steeply falling $pp$ neutrino spectrum with
the falling edge of the RSFP resonance pit.  As
can be seen in Fig.~\ref{fig:3}, the 13.6 y$^{-1}$ frequency associated
with these neutrinos increased in strength from the start of data taking after
the solar maximum of 1989.6
to the solar minimum of 1996.8, after which the modulation becomes weak.  Also it was during
that cycle that the main buildup in the strength of the 12.9 y$^{-1}$
oscillation was detected by Homestake.  The SXT X-ray data, with which these two
frequencies had remarkable agreement \cite{ref:8}, also came from that same
solar cycle.

\begin{figure}
\epsfig{file=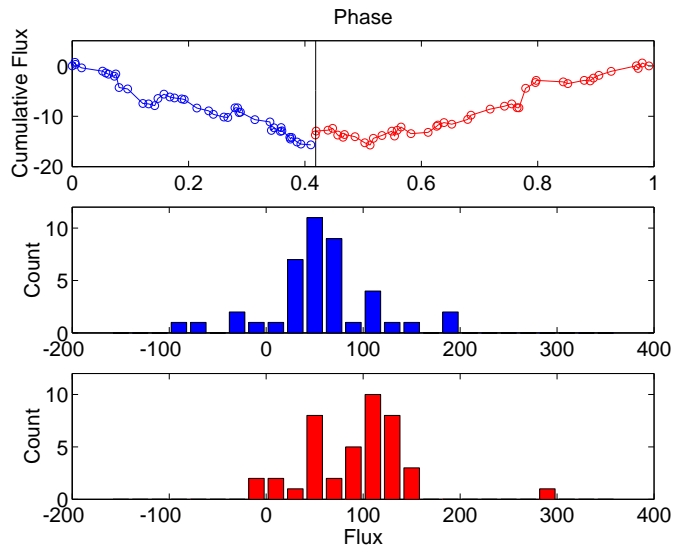,width=3.5in}
%\vspace{2.5in}
\caption{Normalized GALLEX flux measurement runs (prior to 1997) reordered according to the phase
of the 13.59 y$^{-1}$ solar modulation.  The division in phase is made so as to have equal numbers of events in the 
descending (blue) and ascending (red) parts of the cycle.
\label{fig:4}}
\end{figure}

Another feature attributable to a time variation of the intersection of the
$pp$ spectrum with the edge of the RSFP resonance pit is the appearance of a
bimodal flux distribution in the Ga data.  This distribution is clearly evident
\cite{ref:6} during the same solar maximum to solar minimum
period when the flux is binned appropriately
using individual runs, instead of averages over several data runs.  This
neutrino flux effect also diminishes after the solar minimum.
Adding to the significance of this result is the plot of Fig.~\ref{fig:4},
which shows that when the end times of runs are reordered according to the
phase of the 13.59 y$^{-1}$ rotation, the flux values are low in one-half
of the cycle and high in the other.

It is unfortunate that the Homestake experiment, which detected mainly the $^8$B neutrinos (the only neutrino
component registered by Super-Kamiokande) stopped operating at about the time that Super-Kamiokande started.
As a result, there is no way to predict from results of other experiments what neutrino
flux variation should have been detected by Super-Kamiokande during its operation from May 1996
(near solar minimum) until July 2001 (near solar maximum).

Recently the Super-Kamiokande group released \cite{ref:12} flux measurements in 184 bins of about 10 days each. While 
these measurements vary by $\sim 2$, averaged over all bins their fractional error is $0.14$. Because of the regularity of the binning, the ``window spectrum''
(the power spectrum of the acquisition
times) has a huge peak (power $S>120$) at a frequency of $\nu=35.98$ y$^{-1}$
(period 10.15 d).  (Note that the probability of obtaining a power of
strength $S$ or more by chance at a specified frequency is $e^{-S}$.)  This 
regularity in binning leads to aliasing of the power
spectrum of the flux measurements.  In a spectrum formed by a likelihood
procedure, the strongest peak in the range 0 to 100 y$^{-1}$ occurs at $\nu=26.57$ y$^{-1}$
\cite{ref:13a,ref:25} with $S=11.26$.  In the range 0--40 y$^{-1}$,
the next strongest peak is at $\nu=9.41$ y$^{-1}$ with $S=7.33$.  Since
$26.57+9.41=35.98$, we infer that the weaker peak (9.41 y$^{-1}$) is an alias of the
stronger (26.57 y$^{-1}$).  

\begin{figure}
\epsfig{file=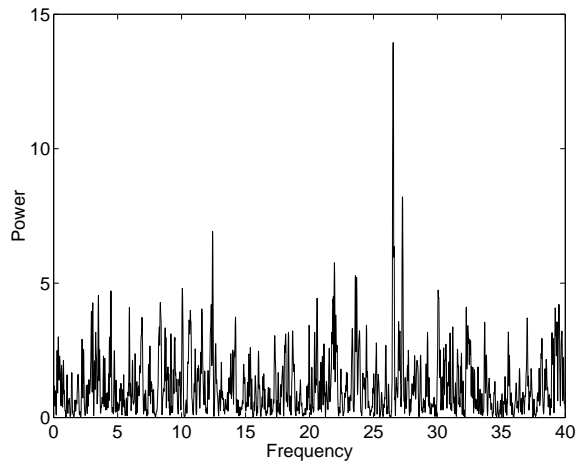,width=3.0in}
\caption{Combined power spectrum of the Super-Kamiokande, SAGE, and GALLEX-GNO data.
\label{fig:5}}
\end{figure}

Having found this frequency in the high-statistics Super-Kamiokande data, we then
examined the power spectrum from a combined analysis of the SAGE \cite{ref:26} and GALLEX-GNO
data, the sampling time of the Homestake data being too long for this high a
frequency.  The combined experiments showed a peak at 26.54$\pm 0.05$ y$^{-1}$ with $S=5.75$.
When combined with the Super-Kamiokande data, we find a peak at 26.54$\pm 0.03$ y$^{-1}$ with $S=13.95$, as shown in
Fig.~\ref{fig:5}.  There is no significant peak near 9.41 y$^{-1}$, further
showing this to be an alias frequency.

Since 26.5 y$^{-1}$ is within the band of twice
the near-equatorial rotation frequencies in the solar interior, this result
indicates that some $^8$B neutrinos are experiencing an ``$m=2$" structure, two
circumferential regions of the magnetic field which differ in strength from
the rest.  This modulation, unlike the Ga $pp$ threshold
effect, is not very deep ($\sim$ 10\%).  Such oscillations at the
harmonic of the rotation frequency are not uncommon in solar data.

%  We also find
%similar but weaker peaks in power spectra formed from both GALLEX-GNO and
%SAGE data.  GALLEX-GNO and SAGE power spectra may be merged into a single
%statistic that has the same distribution, $e^{-S}$, as the power spectrum
%formed from normally distributed random noise with variance unity.  This
%statistic may then be merged, by the same procedure, with the
%Super-Kamiokande power spectrum.  In this way, we are able to display and
%quantify peaks that are common to GALLEX-GNO, SAGE, and Super-Kamiokande.
%In the range $\nu=0$--40 y$^{-1}$, we find one major peak at 26.57
%y$^{-1}$ \cite{ref:26} with equivalent power 12.46.  There is no other
%peak in this range with power in excess of 5.64, and there is no
%significant peak near 9.41 y$^{-1}$, further identifying this frequency
%as an alias of 26.57 y$^{-1}$.

The analysis of Super-Kamiokande data, when combined with the results of
analyses of Cl and Ga data, yields evidence for variability of the solar
neutrino flux.  The KamLAND experiment supports the LMA solution, but it
could be combined with RSFP.  A definitive result on neutrino flux
time dependence can be carried out using a combined analysis from all
experiments, particularly using one-day bins
from all water Cerenkov detectors.

We are indebted to many friends for assistance and helpful discussions.
D.O.C.\ is supported by a grant from the Department of Energy
No.~DE-FG03-91ER40618, and P.A.S.\ is supported by
grant No.\ AST-0097128 from the National Science Foundation.

% now the references. delete or change fake bibitem. delete next three
%   lines and directly read in your .bbl file if you use bibtex.
\gdef\journal#1, #2, #3, #4#5#6#7{      % Journal reference.  Comma sets
    {\sl #1~}{\bf #2}, #3 (#4#5#6#7)}   % off: name, vol, page, year
\def\apj{\journal Astrophys.\ J., }
\def\app{\journal Astropart.\ Phys., }
\def\baas{\journal Bull.\ Am.\ Astron.\ Soc., }
\def\nature{\journal Nature, }
\def\nc{\journal Nuovo Cimento, }
\def\np{\journal Nucl.\ Phys., }
\def\npps{\journal Nucl.\ Phys.\ (Proc.\ Suppl.), }
\def\pl{\journal Phys.\ Lett., }
\def\pr{\journal Phys.\ Rev., }
\def\prc{\journal Phys.\ Rev.\ C, }
\def\prd{\journal Phys.\ Rev.\ D, }
\def\prl{\journal Phys.\ Rev.\ Lett., }
\def\sjnp{\journal Sov.\ J.\ Nucl.\ Phys., }
\def\solarphys{\journal Solar Phys., }
\def\jetp{\journal J.\ Exp.\ Theor.\ Phys., }

\end{document}